\documentclass[iop,jphysa,superscriptaddress,longbibliography,
eprint,
]{revtex4-2}
\usepackage[utf8]{inputenc}
\usepackage[T1]{fontenc}
\usepackage{amsmath,amssymb,amsfonts,amsthm,mathtools,mleftright}
\usepackage{dsfont}
\usepackage{graphicx,grffile,subfigure}
\usepackage[dvipsnames]{xcolor}
\usepackage{url,hyperref}
\hypersetup{colorlinks=true, linkcolor=BrickRed, urlcolor=blue!50!black, citecolor=blue!50!black}
\usepackage{cleveref}
\usepackage{bigints}

\newcommand\FUB{\affiliation{Freie Universität Berlin, Department of Mathematics and Computer Science, Berlin, Germany}}
\newcommand\ZIB{\affiliation{Zuse Institute Berlin, Germany}}

\newcommand{\VEC}[1]{\boldsymbol{#1}}

\newcommand{\eq}[1]{(\ref{#1})}
\newcommand{\reals}{\mathbb{R}}

\newcommand{\VXi}{\VEC{\Xi}}
\newcommand{\VX}{\VEC{X}}
\newcommand{\VY}{\VEC{Y}}
\newcommand{\Vx}{\VEC{x}}

\newcommand{\Vz}{\VEC{z}}

\newcommand{\vF}{\vec{F}}
\newcommand{\vn}{\vec{n}}
\newcommand{\vp}{\vec{p}}
\newcommand{\Vp}{\VEC{p}}
\newcommand{\vq}{\vec{q}}
\newcommand{\Vq}{\VEC{q}}

\newcommand{\Vv}{\VEC{v}}

\newcommand{\Vtotal}{V_{\text{tot}}}
\newcommand{\Fmean}{\vec{F}_{\text{av}}}

\newcommand{\dss}{\displaystyle}

\begin{document}
	
\title{Dynamics of systems with varying number of particles: from Liouville equations to general master equations for open systems}

\author{Mauricio J. del Razo}\email{m.delrazo@fu-berlin.de}
\FUB
\ZIB

\author{Luigi Delle Site}\email{luigi.dellesite@fu-berlin.de}
\FUB

\date{\today}

\begin{abstract}
	A varying number of particles is one of the most relevant characteristics of systems of interest in nature and technology, ranging from the exchange of energy and matter with the surrounding environment to the change of particle number through internal dynamics such as reactions.
	The physico-mathematical modeling of these systems is extremely challenging, with the major difficulty being the time dependence of the number of degrees of freedom and the additional constraint that the increment or reduction of the number and species of particles must not violate basic physical laws. Theoretical models, in such a case, represent the key tool for the design of computational strategies for numerical studies that deliver trustful results.
	In this manuscript, we review complementary physico-mathematical approaches of varying number of particles inspired by rather different specific numerical goals. As a result of the analysis on the underlying common structure of these models, we propose a unifying master equation for general dynamical systems with varying number of particles. This equation embeds all the previous models and can potentially model a much larger range of complex systems, ranging from molecular to social agent-based dynamics.
\end{abstract}

\maketitle

\section{Introduction}
The conception of many-particle (or equivalently, many-body) systems has played a central role in modern theoretical and mathematical physics, from the many-electron systems of quantum mechanics to the tracers of fluid dynamics. Many-particle models are ubiquitous in condensed matter and provide the paradigm of reference for studying the properties of any existing substance. It is also an interesting approach per se in assessing the cascade of scales that characterize the physics of matter, since, in principle, it allows to systematically pass from the microscopic quantum mechanical resolution, up to the continuum hydrodynamics \cite{kuzemsky}.  The models used in physics are in general based on an interaction potential, usually of a two-body nature, (e.g. electrostatic, Lennard-Jones) and the corresponding interlinked Newton's equations of motion (or Schr\"{o}dinger equations in quantum mechanics). This means that particle systems can be mathematically described as dynamical systems moving along calculable trajectories, enabling powerful mathematical and physical machinery to model, simulate and analyze them.

Most of the machinery developed to model and simulate molecular or particle-based dynamical systems focuses on systems with a constant number of particles, this means they are closed but not necessarily isolated from heat exchange. However, this is not the case in many real-world applications, where we often have to focus on subsystems that exchange material with their surroundings; we refer to such systems as open systems. For instance, living cells constantly exchange molecules and energy with their environment; they consume chemical energy and dissipate heat. In terms of physical chemistry, every living system must be a nonequilibrium open system ---a closed system has no life \cite{qian2007phosphorylation}. These nonequilibrium processes at the molecular scale often drive fundamental phenomena such as symmetry breaking, phase transitions and entropy production with profound impact on our understanding of living systems at meso- and macroscopic scales \cite{qian2016entropy, qian2016framework, netz2020approach}.  Moreover, the exchange of heat and matter is one of the main processes driving weather and climate phenomena, such as tipping events \cite{wunderling2024climate}. This manifests in the need for multiscale models \cite{achatz2023multi}, as well as reduced/coarse-grained models \cite{majda2009normal} and stochastic closure techniques \cite{berner2017stochastic}. Thus, to understand, model and simulate these processes, it is fundamental to develop mathematical and physical machinery to handle systems with varying particle number.

From a mathematical perspective, it is a complex problem. The main difficulty is the on-the-fly change in the number of degrees of freedom of the system. At a differential equation level, the number of equations would change as the system changes particle number. The solution is to lift the dynamics to the space of densities/distributions, where the dynamics become linear, albeit infinite-dimensional and thus complicated to analyze. This is analogous to switching representation from a nonlinear dynamical system to the linear Liouville equation, or in the case of stochastic dynamical systems to a Fokker-Planck equation. However, note some information is lost in the process, as we can no longer track individual particles. Following this reasoning, some theoretical machinery has been developed in previous works to handle systems with varying particle number. We will overview the main ones in this manuscript. Other similar approaches often employ methods based on quantum field theory \cite{doi1976second,grassberger1980fock, del2022probabilistic,del2024field}, which are not a requirement to neither write nor understand the equations. From a computational perspective, one possible solution would be to consider a large enough closed subsystem that includes a substantial part of the environment. In this context, numerical methods of particle simulation have enormously risen in the past decades, improving our capability of performing precise calculations of large many-particle systems (see e.g. \cite{frenkel, binder}). However, despite the growing computational power, many systems of major interest are not yet affordable with standard available models, so we need alternative solutions that combine novel mathematical developments with computational approaches.

In recent years, multiscale models and simulations have been developed to improve the efficiency of particle-based simulations. The basic idea is that one retains the degrees of freedom strictly required by the problem and simplifies the degrees of freedom that are not directly involved in the process of interest at coarser levels \cite{annurev}. A prototype situation in most of the many-particle models is the necessity of coarse-graining the environment around the region of interest while retaining all the necessary details in the region itself; since the region of interest is subject to particle number fluctuations, the necessity of developing open many-particle systems rises naturally \cite{epjb}. Moreover, once a mathematical model for the dynamics with varying number of particles is developed, it can organically be adapted and applied to systems where the particle number fluctuations are provided not only by the external environment but also by the change of composition of a system, when for example in a mixture different species interact forming a third one. Despite the high dimensionality of the resulting mathematical model, it can be used as a guiding framework to unify models at multiple scales and to derive physically consistent multiscale numerical schemes.

The concept of subsystem is closely related to the concept of subdynamics for which there is a vast literature available. For example, the model proposed by Prigogine and coworkers \cite{prigo1,prigo2}, the model of Emch and Sewell \cite{emsew}, and the model of Robertson\cite{rob,seke}, to cite but a few. The idea is based on the projection method of Zwanzig \cite{zwanzig} for quantum systems (used also for classical systems). The evolution of the probability density of the system in the von Neumann or Liouville equation is projected on a subspace with a reduced number of variables and the rest of the system is formally coarse-grained.
The models discussed in this paper share the idea of integrating/coarse-graining the degrees of freedom outside the subsystem, however they add the explicit treatment of the number of particles as a variable of the problem. This characteristic is not explicitly mentioned in the literature cited above, thus leaving the impression that the number of particles of the subsystem is assumed to be fixed. In a system with varying number of particles one should explicitly discuss the normalization of the probability density of the subsystem; this operation is substantially different from the case of a fixed number of particles. In addition, the equations proposed in \cite{prigo1,prigo2,emsew,rob,seke}, while mathematically rigorous and certainly elegant, are characterized by a rather complex structure difficult to implement in modern computational simulation techniques. Instead, for computational implementation, one needs models that are characterized by a certain level of mathematical rigor but at the same time are constructed to efficiently capture the relevant physics. The model of Bergman and Lebowitz \cite{leb1,leb2}, discussed later on, can be considered the first historical attempt towards such an effort and became a source of inspiration for most of the progress reported here.

In this work, we first review two frameworks to handle classical systems with varying number of particles (motivated by different physical problems and numerical approaches). We introduce these frameworks and investigate the relations between them to show how a general master equation for systems varying number of particles emerges. The two main different points of view correspond to: (i) the approach based on Liouville-like equations of a subsystem embedded in a reservoir where the degrees of freedom are either explicitly integrated out, reducing the effect of the environment on the boundary conditions of the subsystem of interest \cite{jmp,jpa}, or implicitly integrated out into a coupling term \cite{leb1}; and (ii) a master equation where the diffusion process based on the Fokker-Planck operator is coupled with a reaction process that can change the number of particles. In this case, the particle description of the environment is not considered and its effect is modeled empirically {\it a priori} \cite{del2022formulations,del2022probabilistic}. The first approach is inspired by molecular simulations where the microscopic single molecular trajectories with explicit chemical details are accessible and thus the probability distribution function of the phase space of system and environment can be explicitly sampled \cite{tracers}.
This means that a simulation of an open system with the simplified environment obtained analytically by formal integration of its related degrees of freedom should deliver the same result as an equivalent subsystem in a simulation of the entire system; as a consequence, the validity of the theoretical model of the open system can be directly tested numerically (see e.g. \cite{abbasprl}). The second approach is inspired by problems occurring at a larger scale than the microscopic molecular scale, e.g. processes at the scale of biological living cells, where the number of molecules is large enough to render molecular models intractable but not large enough to consider macroscopic approaches that neglect inherent stochastic fluctuations \cite{qian2010chemical}. From a physics perspective, in this approach, molecules are represented as particles undergoing diffusion and chemical reactions are coarse-grained into events that simply change the chemical composition of the system. Thus, the effect of the environment is modeled empirically through the diffusion constant, the reaction rates and the stochastic effects. This model does not only provide a probabilistic model for reaction-diffusion processes \cite{del2022formulations,del2022probabilistic} but also serves as a starting point to derive other models at different scales. This yields a unifying theory that serves as the backdrop to derive numerical schemes that are consistent across multiple scales \cite{kostre2021coupling,del2018grand,del2024open}.

Independently from the original inspiration, both models deal with varying number of particles and thus must retain a common general structure within a unique framework. In this work, we proceed with the formal analysis of the similarities and differences between the two approaches. Based on this comparison, a generalized structure emerges in the form of a unitary equation, from which the previous approaches are special cases. This generalized structure can be used for any dynamical system in an open setting, enabling a broader range of applications within physics and beyond. To finalize, we discuss mathematical perspectives and potential problems of interest in physical and chemical applications both from an analytical and numerical point of view. We further discuss possible applications of the general master equation beyond the fields that initially motivated the equation.

The paper is structured as follows. To motivate the reader, \cref{sec:genMasterEq} focuses on an intuitive overview of the final general master equation for systems with varying particle number, which emerges naturally in \cref{sec:recoveryPrevMod}. However, the equation itself is originally motivated by the specific approaches presented in \cref{sec:LiouvilleEqns,sec:mastereqRD}. More specifically, in \cref{sec:LiouvilleEqns}, we explore two approaches to handle systems with Hamiltonian structure that are in contact with a material reservoir(s). In \cref{sec:mastereqRD}, we show how to write the master equation to describe the probabilistic dynamics of reaction-diffusion processes, and we extend this to systems where the diffusion is governed by Langevin dynamics. In \cref{sec:recoveryPrevMod}, we show how the mathematical similarities between the approaches from \cref{sec:LiouvilleEqns,sec:mastereqRD} inspire formulating the general master equation and how it recovers the previous approaches as special cases. Finally, in \cref{sec:perspectiv}, we discuss the perspectives and future scope of this work in the context of physical modeling, numerical simulations and applications in other fields.

\section{A general master equation for systems with varying number of particles}
\label{sec:genMasterEq}
The general master equation presented in this section was inspired by the frameworks presented in \cref{sec:LiouvilleEqns,sec:mastereqRD}. For educational purposes, we first carry out an intuitive derivation of the equation. We formulate the master equation for systems with varying number of particles starting from a general random dynamical system perspective. This together with \cref{sec:recoveryPrevMod} constitute the main results of the paper. 

We first consider a random dynamical system with fixed dimension $n$, e.g. a system with a fixed number of particles. The dynamics of the trajectories of the systems are given by the system of stochastic differential equations
\begin{align}
	\dot{\Vx} = F(\Vx),
	\label{eq:generalSDE}
\end{align}
where $\Vx$ could represent many things, depending on the application at hand. In physics, it could represent the position of particles or perhaps the positions and velocities of particles. In other fields like social systems, it could represent the position of an agent in opinion space \cite{helfmann2023modelling} or the population of individuals of a certain species \cite{kot2001elements}. The function $F$ incorporates the deterministic drift as well as noise components. One can equivalently write the dynamics of the probability distribution of the system in phase space given by the corresponding Fokker-Planck equation
\begin{align}
	\frac{\partial }{\partial t}f_n = \mathcal{A}_n f_n,
	\label{eq:generalFPE}
\end{align}
with $\mathcal{A}_n$ the infinitesimal generator of the Fokker-Planck equation for the corresponding $n$ particle system, and $f_n:=f_n(t,\Vx)$ the probability density of the system being at $\Vx$ at time $t$, which naturally integrates to one when integrating over the whole phase space. Reasonable boundary conditions for $f_n$ are usually reflective in a bounded domain or vanishing at infinity for a non-bounded domain.

To write a similar description for a system with a variable number of particles, one cannot write it in the form of \cref{eq:generalSDE} since the dimension of the system changes as time evolves. However, we can write something similar to the master equation from \cref{eq:generalFPE}. Consider the family of distributions $f=(f_0,f_1,\dots,f_n,\dots)$, where $f_n:=f_n(t,\Vx^n)$ is the probability density of having $n$ particles (or whatever else we are modeling) at positions or states $\Vx^n=(x_1,\dots,x_n)$ at time $t$. The phase space now has a much more complex structure with continuous and discrete degrees of freedom, so the normalization condition is
\begin{align}
	\sum_{n=0}^\infty \int f_n(t,\Vx^n) d\Vx^n = 1,
\end{align} 
where the integral is over the whole available space. The master equation for the system with varying particle number can then be written as a family of Fokker-Planck equations, one for each $n$-particle level. These are then coupled by operators $\mathcal{Q}_{nm}$ that transfer probability from the $m$-particle state to the $n$ one modeling processes that change the particle number (e.g. interactions with a material reservoir or reactions). These operators form a coupling matrix, which incorporates into the master equation
\begin{align}
	\frac{\partial}{\partial t} 
	\begin{pmatrix}
		f_0 \\
		f_1 \\
		\vdots \\
		f_n \\
		\vdots
	\end{pmatrix}
	=
	\begin{pmatrix}
		\mathcal{A}_0 f_0 \\
		\mathcal{A}_1 f_1 \\
		\vdots \\
		\mathcal{A}_n f_n \\
		\vdots
	\end{pmatrix}
	+
	\begin{pmatrix}
		\mathcal{Q}_{00} & \mathcal{Q}_{01} & \dots & \mathcal{Q}_{0n} & \dots\\
		\mathcal{Q}_{10} & \mathcal{Q}_{11} & \dots & \mathcal{Q}_{1n} & \dots \\
		\vdots  & \vdots && \vdots& \dots \\
		\mathcal{Q}_{n0} & \mathcal{Q}_{n1} & \dots & \mathcal{Q}_{nn} & \dots\\
		\vdots & \vdots & &\vdots & \ddots
	\end{pmatrix}
	\begin{pmatrix}
		f_0 \\
		f_1 \\
		\vdots \\
		f_n \\
		\vdots
	\end{pmatrix}.
	\label{eq:mainMasterEq}
\end{align}
Each row of the equation corresponds to the dynamics of the $n$-particle distribution. If the coupling matrix is removed, one obtains a set of uncoupled Fokker-Planck equations, each for a different particle number. If the space dependence is removed, one recovers a continuous-time Markov chain, where the operators $\mathcal{Q}_{nm}$ become transition rates. This master equation for systems with varying particle number can be written more compactly by focusing on the $n$th component
\begin{align}
	\left(\frac{\partial }{\partial t} - \mathcal{A}_n\right) f_n = \bigg. \mathcal{Q}_n f,
	\label{eq:genMasterEq}
\end{align}
with $\displaystyle \mathcal{Q}_n f=\sum_{m=0}^\infty \mathcal{Q}_{nm} f_m$. Although these infinite sums might seem problematic at first, in practical cases most $\mathcal{Q}_{nm}$ are zero and the sums remain finite, as we will see later. Moreover, in analogy with stochastic matrices in continuous-time Markov chains, we expect each matrix column to sum to zero to enforce the conservation of probability. As in this case, we have a matrix of operators and space dependence, the probability conserving condition translates into the following constraint imposed into the coupling operators 
\begin{align}
	\sum_{n=0}^\infty \int \left(\mathcal{Q}_{nm} \eta_m\right)(\Vx^n)d\Vx^n =0,
	\label{eq:probConservQnm}
\end{align} 
for $\eta_m$ any given test arbitrary probability density for $m$ particles and where the integrals run over the whole available space. Each of these integrals represents the probability flux associated with a transition. By forcing their sum to be zero, we enforce that the net probability flux leaving one state is the same as the one entering another state(s). Thus, enforcing probability conservation. As a final note, there is a connection between this description and hybrid switching diffusions \cite{mao2006stochastic,yin2010hybrid,bressloff2017stochastic,del2021multiscale}. However, in the particle context, hybrid switching diffusions can only account for a change in the state of a particle but cannot incorporate or remove particles.

In the next sections we will review the derivation of master equations of this form written for specific contexts, which have inspired the generalization above that applies to a larger class of systems with varying number of particles \cite{doi1976second,jmp,jpa,del2022probabilistic,del2022formulations,leb2}. In \cref{sec:LiouvilleEqns,sec:mastereqRD}, we will explore specific physical systems that involve varying particle number. Then, in \cref{sec:recoveryPrevMod}, we will show how these specific systems inspired the general master equation (\cref{eq:genMasterEq}) and how they are recovered as particular cases.

\section{Liouville-like equations for classical open systems}
\label{sec:LiouvilleEqns}
The two approaches reported in this section were developed (the first one) or used (the second one) to conceptually frame a numerical method for the molecular simulation of open systems that exchange energy and matter with a large reservoir. The specific numerical code embedded in this theoretical framework is the Adaptive Resolution Simulation (AdResS) \cite{prl12,prx,tracers}; however, any molecular simulation approach that is characterized by the system-reservoir exchange of energy and particles could be framed as well in such models. The models presented here have been used as inspiration for designing and rationalizing the system-reservoir coupling \cite{noneq1,noneq2,abbasprl}. Systems of molecular simulation are in general characterized by an explicit particle-particle Hamiltonian and by its corresponding phase-space probability density. This latter is not known explicitly, however it is statistically sampled either through a single long trajectory or through a collection of short trajectories, each with an initial condition uncorrelated to the initial condition of the others. As a consequence the statistical calculation of physical quantities is done by sampling and averaging the physical quantity of interest along such trajectories \cite{frenkel}. The natural complete framework for the treatment of these systems is the Liouville equation, thus it seemed natural to manipulate the Liouville equation of the total system to obtain an equivalent equation for a subsystem where the surrounding (rest of the total system) has been explicitly integrated out in its particle degrees of freedom or, alternatively, the particle degrees of freedom have been empirically removed by modeling the reservoir as a generic thermodynamic bath.

\subsection{From a large system of $N$ particles to an open subsystem of $n$ particles}
\label{sec:LiouvilleEqns1}
\begin{figure}
	\centering
	\includegraphics[angle=0,width=0.43\textwidth]{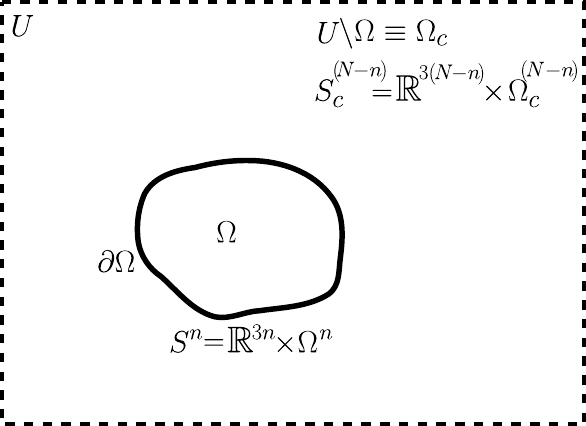}
	\caption{Graphical representation of the open system and associated formalism. The open system $\Omega$, with its boundary surface $\partial\Omega$, is defined as a subsystem of $U$ with $n$ particles. The reservoir is defined as a large system, $U\backslash\Omega=\Omega_{c}$, with $N-n$ particles, since $U$ contains $N$ particles. The $n$ dimensional domain of the phase-space of particles in $\Omega$ is defined as $S^{n}=\reals^{3n}\times \Omega^{n}$ while the $N-n$ dimensional domain of the phase-space of particles in $\Omega_{c}$ is defined as $S_{c}^{N-n}=\reals^{3(N-n)}\times \Omega^{(n-N)}$. This figure is adapted from Fig.1 of \cite{jpa}.} 
	\label{universe}
\end{figure}

The model of open system based on the Liouville equation of the total system of \cite{jmp} is here described in its essential features. Let us consider a large dynamical system of $N$ particles in equilibrium (called here Universe, $U$) and define an open subsystem, $\Omega$, containing $n$ particles (with the corresponding reservoir, $U\backslash\Omega=\Omega_{c}$, of $N-n$ particles, with $N>>N-n$), as illustrated pictorially in \cref{universe}. Starting from the Liouville equations for the probability in phase space of the Universe and integrating out all the degrees of freedom of $U\backslash\Omega=\Omega_{c}$ one would wish to derive a Liouville-like equation for the particles in $\Omega$ taking into account that particles can freely move from the subsystem to the reservoir and vice versa.
The Hamiltonian of the Universe is defined as:
\begin{equation}
	H_{N} = {\sum_{i=1}^{N} \frac{\vp_i^2}{2M} + V_{\rm tot}({\bf q}^{N})}
	\label{uh}
\end{equation}
with $\vp_i$ the momentum of the $i$-th particle, $M$ is the mass of an individual particle; the potential of particle-particle interaction is defined as: 
$V_{\rm tot}({\bf q}^{N})= \sum_{i=1}^N\sum_{j=1, j\not=i}^{N} \frac{1}{2} V(\vq_j-\vq_i)$, with $i$ and $j$ labeling two different molecules with corresponding positions  $\vq_{i}$ and $\vq_{j}$.
The corresponding probability density in phase-space is given by: 
\begin{equation}
	F_N : \dss \reals^+ \times (U \times \reals^3)^N  \to \dss \reals\\ 
	\dss (t, \VX^N)  \mapsto \dss F_N(t,\VX^N)
	\label{probdensu}
\end{equation}
where
\begin{equation}\label{eq:NormalizationUniverse}
	\int_{S^N} F_N\, d\VX^N = 1\,,
\end{equation} 
with {$\VX^N \in S^N = (U \times \reals^3)^N$ the space of the position and momentum variables} of the $N$ particles.\\
The Liouville equation for $U$ is then given by:
\begin{equation}
	\frac{\partial F_N}{\partial t}= {\Lambda}_N F_N
	\label{liuovu}
\end{equation}
with ${\Lambda}_N$ the Liouville operator for an $N$ particle system
\begin{align}
	{\Lambda}_N F_N &= \{F_{N},H_{N}\}=\sum\limits_{i=1}^N 
	\left[
	\nabla_{\vec{q}_i} \cdot \left(\vec{v}_i F_N\right)
	+
	\nabla_{\vec{p}_i} \cdot \left(\vF_i F_N\right)
	\right]\,,
	\label{eq:Liouvillope}
\end{align}
$\{*,*\}$ are the canonical Poisson brackets, $\vec{v}_i = \vec{p}_i/M_i$ is the velocity of the $i$th particle and the total force, $\vF_i$, acting on particle $i$ is given by 
\begin{align}	
	\vF_i &=-\nabla_{q_i}\Vtotal(\Vq^N)= - \sum\limits_{j = 1; j\not=i}^{N}\nabla_{\vq_i} V(\vq_i - \vq_j).
\end{align}

The Hamiltonian of the subsystem $\Omega$ is accordingly defined as:
\begin{equation}
	H_n = \sum_{i=1}^{n} \frac{\vp_i^2}{2M} 
	+ \sum_{i=1}^{n} \sum_{j\not=i}^{n} \frac{1}{2} V(\vq_j-\vq_i) \qquad (\vq_i, \vq_j \in \Omega) ,
	\label{hamom}
\end{equation}
while the corresponding phase-space probability density is:
\begin{eqnarray}
	f_n: \reals^+ \times S^n  \to \reals;~~(t, \VX^n) \mapsto f_n(t, \VX^n) \qquad\text{for}\qquad (n = 0, ..., N)\nonumber\\
	f_n(t, \VX^n) 
	= {{N}\choose{n}}
	\int\limits_{(S_{c})^{N-n}}  
	F_N(t, \VX^n, \VXi_n^{N})\ d\VXi_n^{N}\\
	\VXi_n^{N}  \equiv [\Xi_{n+1},.....\Xi_{N-n}] \qquad\text{where}\qquad
	\Xi_{i}=(\vq_i,\vp_i) \in S_{c}^{{N-n}} . \nonumber
	\label{probdensom}
\end{eqnarray}
The collection of  $n$-particle functions, $(f_n)_{n=0}^N$ defines the probability density with the normalization condition derived from the normalization of $F_{N}$ for the universe (e.g. \eq{eq:NormalizationUniverse}). As a consequence one has that:  
\begin{equation}\label{eq:NormalizationHierarchy}
	\sum\limits_{n = 0}^{N} \
	\int\limits_{\Omega^n} \int\limits_{(\reals^{3})^{n}} f_n(t,(\Vq,\Vp))\ d\Vp\, d\Vq = 1\,.
\end{equation}
The integration of \cref{liuovu} w.r.t.\ the variables of $S_{c}^{N-n}$ implies a straightforward calculation procedure which does not carry any relevant conceptual aspects and thus it is not reported here; the corresponding details of can be found in \cite{jmp,noneq2}, here the final results are reported: 

\begin{equation}
	\frac{\partial f_n}{\partial t} 
	+ {\Lambda}_n f_n
	= \Psi_n + \Phi_{n}^{n+1}
	\label{eqliouvsub}  
\end{equation}
with
\begin{equation} 
	\Psi_n = - \sum\limits_{i=1}^{n} \nabla_{\vec{p}_i}\cdot\Bigl(\Fmean(\vq_i) f_n(t, \VX^{i-1}, X_{i}, \VX_{i}^{n-i})\Bigr)\,,
\end{equation}
and
\begin{equation}
	\Fmean(\vq_i) = - \int\limits_{S^c} \nabla_{\vq_i}V(\vq_i- \vq_j) f^\circ_2(X_j | X_i) d X_j.
\end{equation}
The latter corresponds to the mean field force that the outer particles exert onto the $i$th particle in $\Omega$. The quantity
$f^{\circ}_2(X_{\text{out}} | X_{\text{in}})$ is defined as the conditional distribution of an outer particle at the given state of an inner particle; one can assume that it is known or a model can be proposed. Furthermore,
\begin{equation}\label{eq:BoundaryCondition}
	\Phi_{n}^{n+1}=(n+1)\dss \int\limits_{\partial \Omega} 
	\int\limits_{(\vec{p}_i \cdot \vn) > 0}
	\hspace{-5pt}
	\left(\vec{p}_i \cdot \vn\right) \ 
	\left( {f}_{n+1}\left(t, \VX^{n},  (\vq_i,\vp_i)\right) 
	- {f}_{n}\left(t, \VX^n\right){f}_{1}^{\circ}\left(\vq_i,-\vp_i\right) 
	\rule{0pt}{12pt}\right) 
	d^3 p_i d\sigma_i.
\end{equation}
In \cref{eq:BoundaryCondition} ${f}_{1}^{\circ}\left(\vq_i,-\vp_i\right)$ corresponds to the (modeler's) assumption of the reservoir's one particle distribution calculated at the interface boundary $\partial\Omega$. It must be also added that the addition of an external thermostat in the reservoir does not change the equation and allows us to extend the model to situations of non-equilibrium like thermal gradient \cite{jpa}. In general, such a theoretical framework rationalizes the AdResS molecular dynamics protocol, but in principle, as said above, it can be applied to any other molecular simulation method that aims at considering a varying number of molecules.
\subsection{The Bergman-Lebowitz equation of open systems}
\label{sec:LiouvilleEqns2}
Bergmann and Lebowitz, and Lebowitz and Shimony in two seminal papers \cite{leb1,leb2}, have proposed an equation based on the extension of the Liouville equation that models open systems embedded in a reservoir of particles and energy.
The essence of the model is to consider an impulsive, Markovian interaction between the reservoir and the system; the reservoir is considered stationary and not influenced by the changes occurring in the system, thus the thermodynamic state point of the reservoir is fixed. 
The interaction between the system and the reservoir is modeled as a discontinuous transition of the system from a state with $N$ particles ($\VX^n$) to one with $M$ particles ($\VY^m$), where $\VX^n$ corresponds to the canonical variables (position and momenta). The change of the state of the system state is described by a time-independent Markovian transition kernel, $K_{nm}(\VX^n,\VY^m)$. The kernel,  $K_{nm}(\VX^n,\VY^m)$, corresponds to the probability per unit time that the system {at $\VY^m$ makes a transition to $\VX^n$} as a result of the interaction with the reservoir. 
The probability density, $f_n(\VX^n,t)$, in some point $\VX^n$ of the phase space is then regulated by an extension of the  Liouville equation:
\begin{align}
	\begin{split}
		\frac{\partial f_n(\VX^n,t)}{\partial t}={\Lambda}_{n}f_n+
		+\sum_{m=0}^{\infty}\int d\VY^m[K_{nm}(\VX^n,\VY^m)f_m(\VY^m,t)-K_{mn}(\VY^m,\VX^n)f_n(\VX^n,t)]
	\end{split}
	\label{liouvext}
\end{align}
The first term of \cref{liouvext} corresponds precisely to the Liouville equation for the $n$ particle system (\cref{eq:Liouvillope}), while the second term establishes a coupling with the reservoir(s). This last term is composed of negative or positive contributions corresponding either to the outflux/loss of probability from the current $n$ state or to the influx/gain of probability into the current state, respectively.

If the kernel satisfies the condition of flux balance:
\begin{equation}
	\sum_{m=0}^{\infty}\int [e^{-\beta H(\VX^m){ + \beta \mu m}}K_{nm}(\VX^n,\VY^m)-K_{mn}(\VY^m,\VX^n)e^{-\beta H(\VX^n) { + \beta \mu n}}]d\VX^{m}=0\,,
	\label{gc-cond}
\end{equation}
it follows that the stationary Grand Ensemble is the Grand Canonical ensemble; here $\beta=\frac{1}{k_{B}T}$ where $k_{b}$ is the Boltzmann constant and $T$ the temperature, while $\mu$ is the chemical potential of the system. 
The key difference between the approach of \cref{sec:LiouvilleEqns1} and the approach of Bergman and Lebowitz lies in the assumption about the reservoir and the corresponding term of system-reservoir exchange in the equation. The model of \cref{sec:LiouvilleEqns1} does not require any assumption about the reservoir, but directly integrates its particle's degrees of freedom. As a consequence, the system-reservoir coupling term is explicitly written in terms of particle quantities without any stochastic assumption. In the model shown here, the reservoir is modeled {\it a priori} without any explicit link to its particle resolution and the system-reservoir coupling term is modeled, as a consequence, with a probabilistic process. The model of Bergman and Lebowitz has been very important for the numerical implementation of the open system approach AdResS in an intermediate step. Such a model, differently from the model of \cref{sec:LiouvilleEqns1}, could not provide an explicit numerical receipt for the system-reservoir coupling, nevertheless it allowed for a physical qualitative interpretation of the coupling conditions of the AdResS code \cite{njp,noneq1}.


\section{Master equations for reaction-diffusion processes}
\label{sec:mastereqRD}
\begin{figure}
	\centering
	\textbf{a.}\includegraphics[angle=0,width=0.4\textwidth]{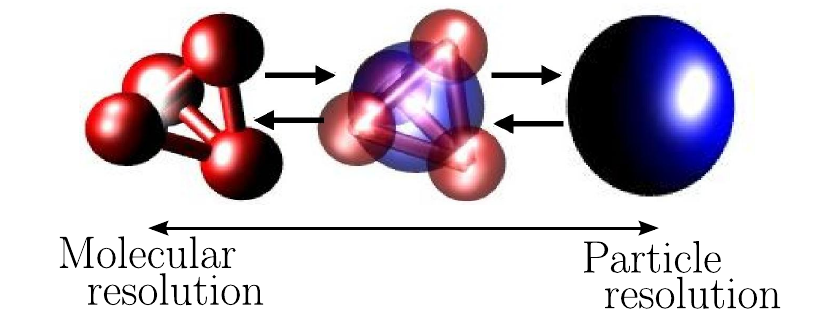} \qquad
	\textbf{b.}\includegraphics[angle=0,width=0.45\textwidth]{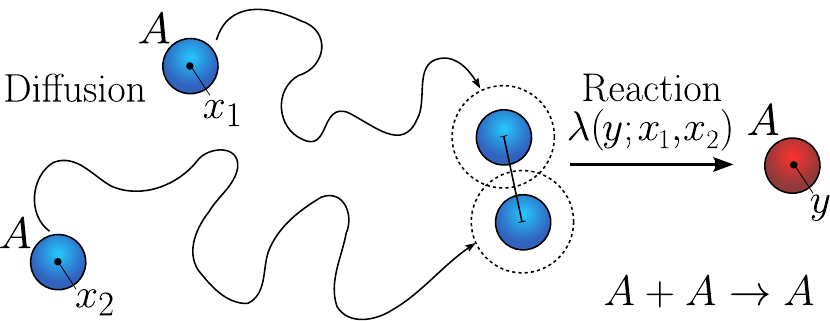}
	\caption{Illustration of particle-based models. \textbf{a.} Change of resolution from molecular to particle resolution, where each molecule is considered as one bead. \textbf{b.} A particle-based reaction-diffusion process for $A+A\rightarrow A$, where $\lambda(y,x_1,x_2)$ is the position dependent rate function. Figure (a) is adapted from Fig.1(a) of \cite{jcpor}.} 
	\label{change-res}
\end{figure}

Molecular dynamics are limited to the study of complex biochemical phenomena at the scale of living cells. While the most sophisticated molecular simulations can achieve simulations of a few macromolecules in the scale of micro- or milliseconds when being optimistic, biochemical processes at these scales often involve thousands or millions of macromolecules and occur over timescales of seconds. Moreover, the chemical events at the molecular scale happen at a much faster scales than those relevant to life processes, and thus their detailed molecular kinetics do not play a key role in the dynamics. It is thus appropriate to consider particle-based reaction-diffusion models (\cref{change-res}), where molecules are represented as particles undergoing random motion due to thermal fluctuations of the solvent (diffusion), and chemical reactions due to instantaneous reaction events, which often occur after chancy encounters between two or more molecules. The relevant features of the molecular scale are captured in the diffusion coefficients and the reaction rate functions. 

Particle-based reaction-diffusion models are often the standard model to describe biochemical processes at the scale of living cells. At these scales, the so-called chemical diffusion master equation (CDME) provides a probabilistic model in terms of the number of particles of each of the chemical species involved, as well as their spatial configuration \cite{del2022probabilistic, del2022formulations,doi1976second}. One of the original objectives of developing the CDME was to serve as an underlying ground model upon which one can construct particle-based reaction-diffusion simulations consistently. Moreover, one can in principle recover most other reaction-diffusion models as limits of the CDME, e.g. in the well-mixed limit, one recovers the well-known chemical master equation (CME) \cite{gardiner1977poisson,gillespie1992rigorous,qian2010chemical,schnoerr2017approximation}; in the thermodynamic limit, one recovers reaction-diffusion PDEs; and when doing spatial discretizations, one recovers the reaction-diffusion master equation \cite{isaacson2009reaction} or the spatiotemporal master equation \cite{winkelmann2016spatiotemporal}. Thus, it serves as a unifying framework from which one can develop numerical schemes that are consistent across scales. For instance, some mathematicians are interested in branching and annihilating Brownian motion \cite{arguin2013extremal} and its connection to the Kolmogorov-Petrovsky-Piskunov-Fisher (KPPF) equation. Following an applied mathematics/mathematical physics approach, one can frame these processes in terms of the CDME and study the limiting cases to recover the KPPF reaction-diffusion PDE, as done recently for other reaction systems \cite{del2024open}. This could yield alternative insights to the research community in that field, as well as methodologies for multiscale simulations.

One of the main virtues of the CDME is its capability to handle systems with varying number of particles while maintaining spatial resolution. This is an inherent characteristic of the CDME since reaction events often change the number of particles in the system. In this section, we will overview the CDME for a simple example, and we will show how it extends when the diffusion term is replaced by Langevin dynamics.  

\subsection{Chemical diffusion master equation: coupling diffusion with reaction processes}
\label{sec:mastereqRD1}

\begin{figure}
	\centering
	\includegraphics[angle=0,width=0.7\textwidth]{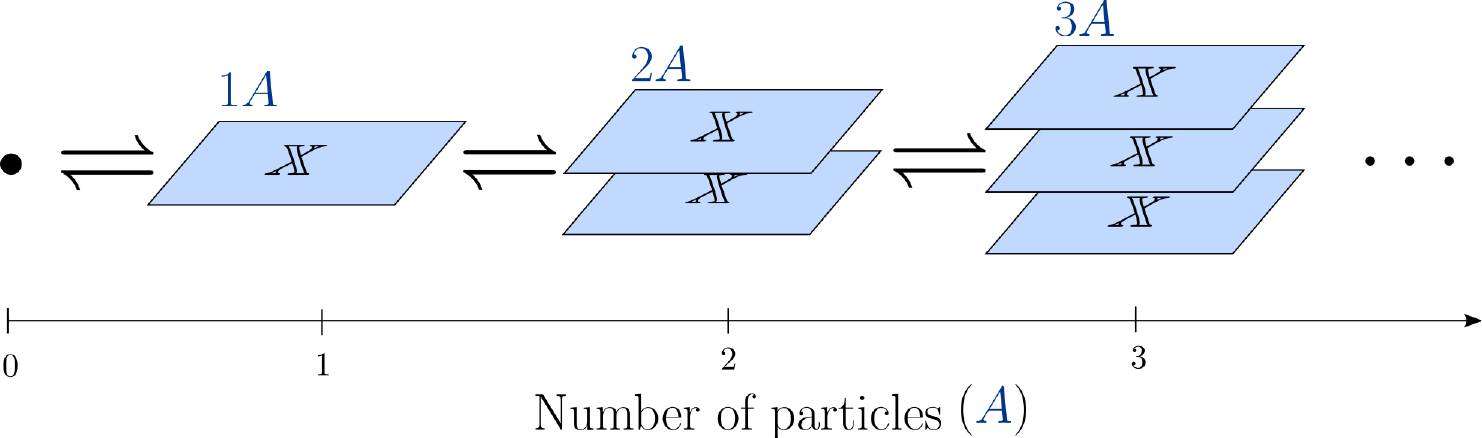}
	\caption{Phase space for a general reaction-diffusion process involving one chemical species. The space $\mathbb{X}$ represents the phase space of one particle $A$ (e.g. $\mathbb{R}^3$), and the distributions described by the chemical diffusion master equation reside in this phase space. Figure adapted from Fig.1(a) of \cite{del2024open}.}
	\label{fig:phasespaceManyA}
\end{figure}

As a starting point, we follow \cite{del2022formulations} and consider a system with a varying number of particles, all corresponding to one chemical species, enclosed in a finite domain $\mathbb{X}$ (with reflecting boundaries in the boundaries of $\mathbb{X}$). The configuration of the system is given by the number of particles and their positions, so its probability distribution is given as an ordered family of probability density functions:
\begin{align}
	f = \left( f_0, f_1, f_2, \dots, f_n, \dots \right),
\end{align}

where $f_n:=f_n(t,\Vq^n)$ is the probability density of finding $n$ particles at the positions ${\Vq^n}=(q_1,\dots, q_n)\in \mathbb{X}^n$ at time $t$, or simply $\Vq$ if clear from context. The phase space of these distributions is depicted in \cref{fig:phasespaceManyA}. As the particles are statistically indistinguishable from each other, the densities must be symmetric with respect to permutations of labels, for instance, $f_3(t,x,y,z) = f_3(t,x,z,y)= f_3(t,y,x,z)$ and so on. The probability distribution should be normalized, thus 
\begin{align} \label{eq:CDMEnormalization}
	\sum_{n=0}^\infty \int_{\mathbb{X}^n} f_n(t,\Vq)d\Vq = 1.
\end{align}
For a system with $M$ reactions, the $n$th component of the CDME has the general form
\begin{equation}
	\frac{\partial f_n}{\partial t} =\mathcal{D}_n f_n + \sum_{r=1}^M \mathcal{R}^{(r)}_n f.
	\label{eq:CDMEgeneral}
\end{equation}
The generator of the CDME is decomposed into the reaction and diffusion components: $\mathcal{D}$ is the diffusion operator and $\mathcal{R}^{(r)}$ corresponds to the reaction operator for the $r$th reaction. Each reaction operator can be split into two contributions 
\begin{align}
	\mathcal{R}^{(r)}_n= \mathcal{G}^{(r)}_n - \mathcal{L}^{(r)}_n,
\end{align}
representing the gain or loss of probability respectively due the $r$th reaction in a given configuration.

To exemplify the formulation of the CDME for a reaction involving only one chemical species, we focus on the following reaction (\cref{change-res}b.)
\begin{equation}
	A+A\rightarrow A.
\end{equation}
The rate at which a reaction event occurs is given by $\lambda(y;\Vx)>0$, and it depends on the positions $\Vx=(x_1,x_2)\in\mathbb{X}^2$ of the reactants and the position $y\in\mathbb{X}$ of the products. Note that the rate function must be symmetric with respect to the positions of the reactants (as well of the products if more than one), i.e. $\lambda(y;x_1,x_2) = \lambda(y;x_2,x_1)$. The $n$th component of the CDME then is given by
\begin{equation}\label{CDME_n}
	\frac{\partial f_n}{\partial t} = \mathcal{D}_n f_n + \mathcal{G}_{n} f_{n+1} - \mathcal{L}_n f_n,
\end{equation}
where $\mathcal{D}_n$, $\mathcal{G}_{n}$, $\mathcal{L}_n$ refer to the corresponding diffusion, gain and loss operators, respectively. Note as we are only considering one reaction, we dropped the upper index. Reactions at the $n$-particle state transfer probability to the $(n-1)$-particle state, so they are represented by the loss term. Reactions at the $(n+1)$-particle state transfer probability to the $n$-particle state state, so they are represented by the gain.

For non-interacting particles, the diffusion operator $\mathcal{D}_n$ is the infinitesimal generator of the $n$-particle Fokker-Planck equation. 
\begin{align}
	\mathcal{D}_n f_n &= \sum_{i=1}^n \nabla_i \cdot \left(A_i f_n\right) + \sum_{i,j=1}^n \nabla_i \cdot \left(D_{i j}\nabla_j f_n\right), 
	\label{eq:diffope}
\end{align}
where $A_i=A_i({\Vq},t)$ is the drift, $D_{ij} = D_{ij}(\Vq)$ are the $3\times 3$ diffusion matrices and $\nabla_i$ denotes differentiation with respect to $i$th  component of the position ${\Vq}$ \cite{del2022probabilistic}. If the drift is consequence of an interaction potential $U({\Vq})$ then
\begin{align}
	A_i = -\sum_{j=1}^n D_{ij}\nabla_j U.
\end{align} 
In the absence of drift and assuming isotropic diffusion, $\mathcal{D}_n$ would simply be the Laplacian, $\mathcal{D}_n=\sum_{i=1}^n D\nabla^2_{i}$ with a scalar diffusion constant $D$. Considering this structure and assuming there are no reactions, the CDME would correspond to an infinite family of uncoupled Fokker--Planck equations for the particle positions, where each member of the family corresponds to a different number of particles. 

The loss operator acting on the $n$-particle density will output the total rate of probability loss of $f_n$ due to all possible combinations of reactants. It is given in terms of the loss per reaction $L_{i, j}$, which acts on $2$ particles at a time, with $(i,j)$ denoting the indexes of the particles that it acts on. The loss per reaction quantifies how much probability is lost to the current state due to one reaction, it is thus the integral over the density and the rate function $\lambda$ over all the possible positions of the products:
\begin{align}
	\left( L_{i,j} f_n \right) ({\Vq}) = f_n({\Vq}) \int_{\mathbb{X}} \lambda(y;\Vq_{i,j}) dy,
	\label{eq:lossperreaction}
\end{align}
with $\Vq\in \mathbb{X}^n$ and where $\Vq_{i,j}=(q_i,q_j)$ represent the $i$th and $j$th components of $\Vq$. The total loss is then the sum of the loss per reaction over all possible reactions,
\begin{align}
	\mathcal{L}_n= \sum_{1\leq i < j \leq n}L_{i,j}.
	\label{eq:loss2A-1A}
\end{align}

The form of the ordered sum guarantees that we count all the possible ways of picking up $k$ particles without double counting. Similarly, the gain operator acting on the $n$-particle density will output the total rate of probability gain of $f_n$. It can be expressed in terms of the gain per reaction resulting from $2$ reacting particles producing one product with index $k$. The gain per reaction, termed $G_{k}$, quantifies how much probability is gained by the current state due to one reaction, it is thus the integral over the density and the rate function $\lambda$ over all the possible positions of the reactants:
\begin{align}
	\left(G_{k} \rho_{n+1} \right) ({\Vq}) = \int_{\mathbb{X}^2}
	\lambda(q_{k};\Vz)
	\rho_{n+1}({\Vq}_{ \setminus \{k \}}, \Vz)  d \Vz,
	\label{eq:gainperreaction}
\end{align}
with $\Vq\in \mathbb{X}^n$ and
where the subscript $\setminus \{k\}$ means that the entry with index $k$ is excluded from the tuple ${\Vq}$ of particle positions. The total gain is then the sum of the gain per reaction over all possible reactions,

\begin{subequations} \label{eq:gainkA-lA}
	\begin{align}
		\mathcal{G}_n & =   \frac{n(n+1)}{2} \frac{1}{n} \ \
		\sum_{k=1}^{n} G_{k},  \label{eq:gain2A-1A1}  \\
		& =   \frac{(n+1)}{2} \ \
		\sum_{k=1}^{n} G_{k}
	\end{align}
\end{subequations}
where in the first line, the first fraction represents all the possible ways of picking two particles of the same species from state $n+1$, and the second fraction and sum represent all the ways of adding a particle into the $n$ state (while preserving symmetry). Gathering all the terms, the CDME then has the form
\begin{align*}
	\frac{\partial f_n(\Vq)}{\partial t} = \sum_{\nu=1}^n D_\nu f_n + \frac{(n+1)}{2} \ \
	\sum_{k=1}^{n}  \int_{\mathbb{X}^2}
	\lambda(q_{k};\Vz)
	f_{n+1}({\Vq}_{\setminus\{k \}}, \Vz)  d \Vz -  \sum_{1\leq i < j \leq n} f_n({\Vq}) \int_{\mathbb{X}} \lambda(y;\Vq_{i,j}) dy,
\end{align*}
where the notation omitted the time dependence of $f_n$ for simplicity. The extension to multiple species is reported in the following subsection. The CDME for general one-species and multiple-species reactions were formulated in detail in \cite{del2022formulations,del2022probabilistic}.

\subsubsection{Multiple species extension}\label{sec:ABC}
To exemplify the CDME for a reaction involving multiple species, we follow  \cite{del2022formulations} again and consider the reaction 
\begin{equation}
	A+B \rightarrow C
\end{equation}
with rate function $\lambda(y; x_A,x_B)$, where $x_A$ and $x_B$ are the locations of one pair of reactants and $y$ is the location of the product. The stochastic dynamics of the system are described in terms of the distributions $f_{a,b,c}\left(\Vq^{a},\Vq^{b},\Vq^{c}\right)$, where $a,b,c$ indicate the numbers of $A$, $B$, and $C$ particles, respectively, and $\Vq^{a}$ indicates the positions of the $A$ particles, $\Vq^{b}$ of the $B$ particles, and $\Vq^{c}$ of the $C$ particles.  The normalization condition \cref{eq:CDMEnormalization} generalizes to multiple species
\begin{align}
	\sum_{a,b,c=0}^\infty\; \int_{\mathbb{X}^a\times\mathbb{X}^b\times\mathbb{X}^c}\ f_{a,b,c}\left(\Vq^{a},\Vq^{b},\Vq^{c}\right) d\Vq^{a}\,d\Vq^{b}\,d\Vq^{c} = 1.
\end{align}
Applying analogous reasoning as before, we can derive the CDME for the bimolecular reaction \cite{del2022formulations}. Here we just state the final equation
\begin{align}\begin{split}\label{CDME:ABC}
		\frac{\partial f_{a,b,c}}{\partial t} = & \sum_{\mu=1}^a D^A_{\mu}f_{a,b,c} + \sum_{\nu=1}^b D^B_{\nu}f_{a,b,c} +\sum_{\xi=1}^c D^C_{\xi}f_{a,b,c}  \\
		& +\frac{(a+1)(b+1)}{c}\sum_{\xi=1}^{c}  \int_{\mathbb{X}^2}	\lambda\left(q^{c}_\xi; z,z'\right) f_{a+1,b+1,c-1}
		\left( (\Vq^{a},z),(\Vq^{b},z'),\Vq^{c}_{\setminus \{\xi\}} \right)
		dz\,dz' \\
		&- f_{a,b,c}\sum_{\mu=1}^a \sum_{\nu=1}^b  \int_{\mathbb{X}} \lambda\left(y; q_{\mu}^{a},q_{\nu}^{b}\right) dy, 
	\end{split}
\end{align}
where to simplify notation, the dependence of $\rho_{a,b,c}$ on time $t$ has been omitted, as well as its dependence on the positions $\left(x^{(a)},x^{(b)},x^{(c)}\right)$ if clear from context. The first line corresponds to the diffusion of each chemical species, the second line to the gain term and the last one to the loss. The works \cite{del2022formulations, del2022probabilistic} developed a comprehensive and formal mathematical framework to formulate the CDME for any given reaction system.

\subsection{Langevin dynamics with varying number of particles}
\label{sec:mastereqRD2}
There is a close mathematical link between \cref{sec:LiouvilleEqns} and \cref{sec:mastereqRD1}; the resulting equations from both approaches describe the dynamics of densities in the phases space of systems with varying number of particles. In \cref{sec:mastereqRD1}, we introduced the diffusion operator, as well as the densities, only dependent on the position of the particles. However, one can in principle define these operators on more general dynamics. For instance, the state of the system can also depend on the velocities of the particles, as well as on the interactions between particles. In \cref{sec:LiouvilleEqns} we had already incorporated velocities, but in a classical context, where there was no stochastic component and the changes in particle number were due to being in contact with a reservoir. A natural question arises: what is the physical/mathematical connection between these approaches?

In this section, we focus on Langevin dynamics with varying copy numbers as a middle ground to hint at the similarities between the classical open systems approach from \cref{sec:LiouvilleEqns} and the stochastic reaction-diffusion approach from \cref{sec:mastereqRD1}. This will hint at a general approach to model the dynamics of systems with varying number of particles, enabling richer models of open systems.

To start, consider the Langevin dynamics of a system with $n$ point particles in a spatial domain $\mathbb{X}$ with positions and velocities $\Vq\in\mathbb{X}^n$ and $\Vv\in\mathbb{V}^n$, where $\mathbb{V}$ is the space of one-particle velocities (in general $\mathbb{R}^3$). The particle's masses are $m$, and they are under an interaction potential $U({\Vq})$ with a configuration-dependent friction tensor $\eta$ to model velocity-dependent hydrodynamic interactions (assuming approximation of pair-wise additivity \cite{ermak1978brownian}), then Langevin dynamics are given by

\begin{align}
	d \Vq(t)=\Vv(t)dt \qquad
	m\cdot d \Vv(t)= -\eta \cdot \Vv(t)dt - \nabla_q U(\Vq)dt + \sqrt{2k_B T} \eta^{1/2} d \VEC{w}(t),
	\label{eq:langevinHydro}
\end{align}
where $\VEC{w}(t)$ corresponds to a $3n$ dimensional vector of independent Wiener processes or standard Brownian motion; $k_B$ is the Boltzmann constant, and $T$ is the temperature.
The corresponding Fokker-Planck equation for these dynamics ---also named Klein-Kramers equation \cite{van1992stochastic}--- determines the dynamics of the probability density $f_n(t,{\Vq},{\Vv})$ in phase space, and it is given by \cite{ermak1978brownian}
\begin{align}
	\frac{\partial f_n}{\partial t}&= \mathcal{K}_n f_n \\
	\text{with} \qquad
	\mathcal{K}_n &=\sum_{i=1}^n \left( -v_i \cdot \nabla_{q_i} f_n + \frac{1}{m} \nabla_{q_i} U \cdot \nabla_{v_i} f_n \right)
	+\sum_{i,j=1}^n \frac{1}{m}\nabla_{v_i} \cdot \ \eta_{ij} \left(
	v_j f_n + \frac{k_B T}{m} \nabla_{v_j}f_n \right)    \label{eq:Lrhon} 
\end{align}
where $\mathcal{K}_n$ simply denotes the infinitesimal generator of this Fokker-Planck equation for an $n$ particle system with positions and velocities---in three dimensions, the equation would be a Fokker-Planck equation in $6n$ dimensions. We would like to now incorporate processes that change the number of particles. This can be framed quite generally following \cref{sec:mastereqRD1} in terms of reactions. Reaction events can be thought of more generally as instantaneous events that follow a Poisson process in terms of rate functions $\lambda$. Note these rates must not be constant; they can depend on time, on the state of the system or on external variables. The master equation for this process will have the same form as the CDME in \cref{eq:CDMEgeneral}, but now instead of the diffusion operator, we have the Klein-Kramers one
\begin{align}
	\frac{\partial f_n}{\partial t} = \mathcal{K}_n f_n + \sum_{r=1}^M\mathcal{R}_n^{(r)} f.
	\label{eq:CDMEhydrogeneral}
\end{align}
Here $M$ denotes the number of processes changing the number of particles, and we assume that $f=(f_0,f_1,\dots,f_n,\dots)$  now also depends on the velocities. Note we again assume the densities are symmetric with respect to particle indexing permutations and that the normalization from \cref{eq:CDMEnormalization} requires additional integration over the velocity variables
\begin{align} \label{eq:Langevinnormalization}
	\sum_{n=0}^\infty  \int_{\mathbb{X}^n \times \mathbb{V}^{n}}  f_n(\Vq,\Vv)\ d\Vq \ d\Vv= 1,
\end{align}
In addition, the rate functions within the $\mathcal{R}$ operators could also depend explicitly on the velocities of particles, allowing for richer particle-base models and generalizing the depth of field of the CDME. The approach presented here based on Langevin dynamics is an ideal starting ground to couple open particle-based models with fluids and to incorporate hydrodynamic and electrostatic interactions. As a final remark, note that if we take the overdamped limit of the Langevin dynamics, we in principle recover the CDME from \cref{sec:mastereqRD1}.

\section{Previous models as special cases}
\label{sec:recoveryPrevMod}
In this section, we first show what are the mathematical connections between the approaches from \cref{sec:LiouvilleEqns1,sec:LiouvilleEqns2} and the other two approaches in \cref{sec:mastereqRD1,sec:mastereqRD2}. Based on the insight gained in \cref{sec:mastereqRD2}, we compare to the previous results and show how this inspired the general master equation for classical systems with varying number of particles from \cref{sec:genMasterEq}. 

In both previous \cref{sec:LiouvilleEqns,sec:mastereqRD}, we introduced equations to describe the dynamics of densities in the phase space of systems with varying number of particles. The first insight is that mathematically the $n$-particle Liouville equation is a special case of the $n$ particle Fokker-Planck equation (Klein-Kramers eq.). To show this, we start with the classical approach from \cref{sec:LiouvilleEqns1}. The resulting equation for the distribution dynamics in the phase space of the open system is given by a family of Liouville $n$-particle equations coupled by the terms $\Psi_n + \Phi_{n}^{n+1}$ (\cref{eqliouvsub})
\begin{equation}
	\frac{\partial f_n}{\partial t} 
	+ \sum\limits_{i=1}^{n} \left(\nabla_{\vq_i}\cdot\left(\vec{v}_i f_n\right)  
	+ \nabla_{\vp_i}  \cdot \left(\vF_i f_n\right)  \rule{0pt}{12pt}\right)
	= \Psi_n + \Phi_{n}^{n+1}.
	\label{eqliouvsub2}
\end{equation}
The Bergman-Lebowitz model from \cref{sec:LiouvilleEqns2} shares the same mathematical structure but with different coupling terms due to different conceptual starting points of view: \cref{sec:LiouvilleEqns1} employs an explicit particle reservoir whose coupling with the open system has been achieved by explicit integration of its degrees of freedom, while \cref{sec:LiouvilleEqns2} employs a reservoir modeled a priori with a stochastic term for the exchange with the open system. In practice, the Bergman-Lebowitz model conceptually lies in between the other two approaches shown here.
We can compare this equation with the $n$th component of the master equation \cref{eq:CDMEhydrogeneral} from \cref{sec:mastereqRD2}
\begin{align}
	\frac{\partial f_n}{\partial t} &- \mathcal{K}_n f_n = \sum_{r=1}^M\mathcal{R}^{(r)}_n f
	\label{eq:CDMEhydrogeneral2}
\end{align}
with
\begin{align}
	\mathcal{K}_n f_n &=  \sum_{i=1}^n \left( -v_i \cdot \nabla_{q_i} f_n + \frac{1}{m} \nabla_{q_i} U \cdot \nabla_{v_i} f_n \right)
	+\sum_{i,j=1}^n \frac{1}{m}\nabla_{v_i} \cdot \ \eta_{ij} \left(
	v_j f_n + \frac{k_B T}{m} \nabla_{v_j}f_n \right).
\end{align}
We can rewrite the first term of the diffusion operator in exactly the same form as \cref{eqliouvsub2}
\begin{align}
	\mathcal{K}_n f_n &=  -\sum_{i=1}^n \left( \nabla_{q_i} (v_i \cdot f_n) + \frac{1}{m} \nabla_{v_i} \cdot \left(F_i f_n \right) \right)
	+\sum_{i,j=1}^n \frac{1}{m}\nabla_{v_i} \cdot \ \eta_{ij} \left(
	v_j f_n + \frac{k_B T}{m} \nabla_{v_j} f_n \right),
	\label{eq:DnLangevin}
\end{align}
where $F_i$ is the net force acting on particle $i$ due to potential-based interactions and $\nabla_{p_i} = \nabla{v_i}/m$. If the noise term of the Langevin equation is removed, the second term of $\mathcal{K}_n f_n$ vanishes recovering exactly the Liouville equation. Thus, the Liouville equation can be mathematically understood as a special case of the Fokker-Planck equation for Langevin dynamics in the deterministic limit. Note the emphasis on ``mathematically'' since from a physical perspective Langevin dynamics are often understood as a coarse-grained representation of classical molecular dynamics.

The second insight is that these equations have an analogous mathematical structure. One part is essentially a transport term for the $n$-particle density given by a Fokker-Planck equation, which simplifies to a Liouville equation in deterministic cases. The other part is a coupling term that models the change in the number of particles across the family of densities $f=(f_0,f_1,\dots,f_n,\dots)$. We can write both equations in the form of \cref{eq:genMasterEq}:
\begin{align}
	\underbrace{\left(\frac{\partial }{\partial t} - \mathcal{A}_n\right) f_n}_{\substack{n\text{-particle}\\ \text{transport term}}} = \underbrace{\bigg. \mathcal{Q}_n f}_{\substack{\text{coupling}\\ \text{term}}}.
	\label{eq:generalstruct}
\end{align}

We can recover all models investigated in the previous sections as follows:
\begin{itemize}
	\item \Cref{sec:LiouvilleEqns1}, Liouville-like equation for an open subsystem:
	\begin{align}
		\mathcal{A}_n f_n \rightarrow {\Lambda}_n f_n \qquad \text{and}\qquad \mathcal{Q}_nf \rightarrow \Psi_n + \Phi_{n}^{n+1}.
	\end{align}	
	In this case, the dynamics of the system are deterministic with Hamiltonian structure, so we only need the Liouville operator for the transport part. All the physics for the coupling between the subsystem with $n$ particles and the larger system (universe) with $N$ particles are condensed in the coupling term. As an additional verification of physical consistency, it was further shown that in equilibrium and under the hypothesis of short-range interactions, this approach recovers automatically the standard stationary grand canonical distribution \cite{jmp}.
	
	\item \Cref{sec:LiouvilleEqns2}, Bergman-Lebowitz equation of open systems:
	\begin{align}
		\mathcal{A}_n f_n \rightarrow {\Lambda}_n f_n \qquad \text{and}\qquad \mathcal{Q}_nf \rightarrow \sum_{m=0}^{\infty}&\int d\VY^m[K_{nm}(\VX^n,\VY^m)f_m(\VY^m,t)-K_{mn}(\VY^m,\VX^n)f_n(\VX^n,t)].
	\end{align}	
	Once again the dynamics of the transport process are deterministic with Hamiltonian structure, so we can use again the Liouville operator. However, the interaction with the reservoir is modeled stochastically using the Markovian transition kernel $K_{mn}$. This can be understood under the same light as the reaction in \cref{sec:mastereqRD1}. We can even separate the terms into a general loss and a gain term $     \mathcal{Q}_n = \mathcal{G}_n - \mathcal{L}_n$ where 
	\begin{align}
		\mathcal{L}_nf &= \sum_{m=0}^{\infty}\int d\VY^m[K_{mn}(\VY^m,\VX^n)f_n(\VX^n,t)] \label{eq:losstranskernel}\\ 
		\mathcal{G}_n f &= \sum_{m=0}^{\infty}\int d\VY^m[K_{nm}(\VX^n,\VY^m)f_m(\VY^m,t)] 
		\label{eq:gaintranskernel}
	\end{align}	
	In this sense, this approach is a middle ground between the approach in \cref{sec:LiouvilleEqns1} and the one in \cref{sec:mastereqRD1}.
	
	\item \Cref{sec:mastereqRD1}, Chemical diffusion master equation for reaction diffusion:
	\begin{align}
		\mathcal{A}_n f_n \rightarrow \mathcal{D}_n f_n \qquad \text{and}\qquad \mathcal{Q}_nf \rightarrow \sum_{r=1}^M \mathcal{R}_n^{(r)}f.
	\end{align}
	In this case, the transport dynamics are stochastic, but they correspond to Brownian dynamics (overdamped Langevin dynamics to be precise), so there is no velocity dependence. Thus the transport part is governed by the Fokker-Planck generator for the standard Brownian diffusion of $n$-particles: $\mathcal{D}_n$. The reaction operators $\mathcal{R}_n^{(r)}$ transport probability between configurations with different number of particles. These can be separated into the gain and loss parts: $\mathcal{R}^{(r)}_n=\mathcal{G}^{(r)}_n-\mathcal{L}^{(r)}_n$. The total gain and loss due to all the reactions is
	\begin{align}
		\mathcal{G}_n = \sum_{r=1}^M \mathcal{G}_n^{(r)} \qquad \mathcal{L}_n = \sum_{r=1}^M \mathcal{L}_n^{(r)}
	\end{align}
	Following \cite{del2022formulations}, we can use the local rate functions for each reaction with these relations and write a global Markovian transition kernel in the form of \cref{eq:losstranskernel,eq:gaintranskernel}, which establishes a direct connection with the Bergman-Lebowitz approach. The main difference is that, in the Bergman-Lebowitz approach, this kernel can depend on velocities while here it would only depend on positions.
	
	\item \Cref{sec:mastereqRD2}, Langevin dynamics with varying particle number:
	\begin{align}
		\mathcal{A}_n f_n \rightarrow \mathcal{K}_n f_n \qquad \text{and}\qquad \mathcal{Q}_nf \rightarrow \sum_{r=1}^M \mathcal{R}_n^{(r)}f.
	\end{align}
	From a mathematical point of view, this case covers the most general transport dynamics from all the examples presented; all the other cases are special cases of this one. Here, the transport dynamics are stochastic and track positions and velocities. Thus, the transport operator is given by the generator of the Klein-Kramers equation; the Fokker-Planck equation for Langevin dynamics. In the overdamped limit, we recover the diffusion operator $\mathcal{D}_n$ from \cref{sec:mastereqRD1}. Alternatively, if we remove the noise term, we recover the Liouville operator $\mathcal{H}_n$ from \cref{sec:LiouvilleEqns1,sec:LiouvilleEqns2}. The coupling term works exactly as before, with the slight difference that now the reaction rate functions can also depend on the velocity. Following the same reasoning as before, we could once again write this in the form of \cref{eq:losstranskernel,eq:gaintranskernel}, but unlike the overdamped Langevin case, we would also have a dependence on velocities.
\end{itemize}

All the cases are special cases of \cref{eq:generalstruct}. However, its mathematical structure is even more general and does not need to be constrained to Langevin dynamics, Hamiltonian structure or a specific form of the coupling terms. The Liouville/Fokker-Planck term can be written for very general deterministic or stochastic dynamical systems, and the coupling terms can model reactions, interactions with external reservoirs or any other process that changes the number of particles. The only constraint is that every term in the equation must conserve the total probability. We know the Fokker-Planck equation part is probability preserving, so we only need to make sure that the couplings $\mathcal{Q}_n f$ also preserve the total probability ( \cref{eq:NormalizationHierarchy,eq:CDMEnormalization,eq:Langevinnormalization}) following \cref{eq:probConservQnm}. This was shown in \cite{jmp,leb1} for classical molecular dynamics and in \cite{del2022probabilistic} for reaction-diffusion dynamics. For other specific applications, one must construct the corresponding transport and coupling operators.

\section{Perspectives: physical, numerical and beyond}
\label{sec:perspectiv}
The general formulation of the master equation for systems with varying particle number presented in this work provides a novel and synergistic link between different fields and how they handle open settings. Through its general formulation, it unveils new perspectives and opens up the application scope to a diverse range of fields within and beyond physics. There is ample potential for future applications, specifically in the design of multiscale theory and simulation of complex systems. In this section, we discuss new perspectives inspired by this work, as well as current and future applications.

From a physics perspective, understanding the dynamics of systems with varying number of particles requires imposing fundamental physical constraints to \cref{eq:generalstruct}. In most cases, these systems will have ---partly--- a Hamiltonian structure, which is represented by the Liouville part of the equation (\cref{sec:LiouvilleEqns}). Thus, in the spirit of the Bergman-Lebowitz approach, it is illustrative to present an alternative separation of \cref{eq:generalstruct} into the following components:
\begin{align}
	\underbrace{\bigg. \frac{\partial f_n}{\partial t} - {\Lambda}_n f_n}_{\substack{n\text{-particle}\\ \text{Liouville}}} = \underbrace{\bigg. \mathcal{T}_n f_n}_{\substack{\text{heat}\\ \text{exchange}}} + \underbrace{\bigg. \mathcal{Q}_n f}_{\substack{\text{material}\\ \text{exchange}}},
	\label{eq:generalphysstruct}
\end{align}
where the infinitesimal generator in \cref{eq:generalstruct} for the corresponding Fokker-Planck equation is $\mathcal{A}_n=\Lambda_n+\mathcal{T}_n$. The terms represent the Liouvillian dynamics (Hamiltonian); the contribution due to stochastic fluctuations, often a thermostat modeling heat exchange with a reservoir(s); and the coupling terms capturing the material/particle exchange with a reservoir(s). If we assume Langevin dynamics, the thermostat will have the form of \cref{sec:mastereqRD2} and $\mathcal{A}_n\rightarrow \mathcal{K}_n$; however, \cref{eq:generalphysstruct} is not limited to Langevin thermostats. The material exchange can model reactions, interactions with a reservoir or any other process that changes the number of particles. Analyzing and manipulating these components could offer insights into the intricate interplay between system dynamics and its surroundings, shedding light on emergent behaviors, equilibrium states and non-equilibrium phenomena. As an example in a simpler context, it has been shown that, in well-mixed biochemical systems,
reservoir interaction drives phenomena fundamental for life processes such as symmetry breaking, entropy production and phase transitions \cite{qian2010chemical,qian2016entropy, qian2016framework}. The form of \cref{eq:generalphysstruct} hints at the possibility of performing a two-level coarse-graining of a Hamiltonian system. First, coarse-graining the fast-scales, e.g. the solvent dynamics into a thermostat. Then, a second coarse-graining following \cref{sec:LiouvilleEqns1} to model the material exchange of the subsystem with the environment. Although these two effects produced by different types of reservoir interaction were also captured in a seminal work \cite{leb1} with equations in the form of \cref{liouvext}, the approach discussed in this work is more general and conceptually different.

In the context of coarse-graining techniques, the Mori-Zwanzig formalism stands out as a powerful tool for capturing the relevant/slow dynamics of complex systems while accounting for memory effects and non-Markovian behavior. Although it is not trivial how to apply it to systems with varying particle number, it sheds light on what kind of dynamics we should expect in the coarse-grained variables. When applying the Mori-Zwanzig formalism to a Hamiltonian system, it not only yields a noise term representing the thermostat but also a term modeling non-Markovian memory effects. What are the memory effects emerging from the coarse-graining due to particle exchange with the reservoir? Can these memory effects be incorporated into the master equation \cref{eq:generalphysstruct}? Can we incorporate memory effects into reaction-diffusion processes and what is their relevance? These are open questions motivated by this work that are tremendously relevant to the physics community.

The quantum mechanical perspective could also be framed in similar terms, particularly through analysis of frameworks like the Lindblad equation \cite{manzano2020short}, a generalization of the von Neumann's equation that describes the time evolution of density matrices/operators for open quantum systems subject to dissipative processes. Von Neumann's equation is the quantum analog to Liouville's equation, where the classical Poisson bracket is substituted by the commutator and the density by the density matrix/operator. The Lindblad equation corresponds to von Neumann's equation with an additional term to model the interactions with the environment. This term is constructed using the so-called jump operators that can be understood as modeling probability ``jumping'' from one state to another due to interactions with the environment, analogously to the $\mathcal{Q}_n f$ term in \cref{eq:generalphysstruct}. It is natural to suggest that one may be able to construct a framework, following the ideas of this manuscript, similar but alternative to the Lindblad master equation, to describe the exchange of particles between arbitrary quantum systems and the environment. Recently an approach similar to that of \cite{jmp} and applied to the von Neumann equation has been proposed by one of the authors \cite{ana}, while the other author reviewed field theory inspired representations of reaction-diffusion systems \cite{del2024field}, which perhaps aids in extending these ideas to a quantum setting.	Nevertheless, a more precise connection to quantum systems and situations out of equilibrium requires a more detailed analysis beyond the current scope and is left for future work.

From a numerical perspective, multiscale simulations play a pivotal role in understanding complex open systems. By having a master equation for the underlying continuous particle-based process, like the one presented in this work, one has a ground model from which coarser models emerge from the bottom up through discretizations, mean-field limits or moment expansions. In particular, systems in contact with material reservoirs are essential for real-world applications, such as molecular dynamics, biochemical kinetics and weather modeling. Reservoirs in these systems can exhibit dynamic behavior and undergo changes in properties over time. 
Modeling such reservoirs accurately requires sophisticated simulation techniques that account for multiscale phenomena and ensure physical consistency across scales.
The formal derivation of the master equation and the coupling term with accurate statistical properties provides a guiding protocol for designing physically well-founded coupling schemes. Some examples of numerical schemes that handle the coupling with dynamic reservoirs accurately are in models related to this work \cite{annurev,del2018grand, kostre2021coupling,tracers,del2024open}, including thermodynamic aspects \cite{abbas1,del2018grand}.
Moreover, these master equations can be used to derive meso/macroscopic models by calculating expectations, higher moments \cite{isaacson2022mean,kostre2021coupling} or even by applying probabilistic limits, such as the law of large numbers, the central limit theorem or large deviation principles \cite{anderson2015stochastic,del2018grand,isaacson2021reaction}. Through these methodologies, a physical and mathematical consistency is established across scales, often in the form of relations between the parameters of the models used at different scales. Then these relations can be used to develop numerical schemes that are physically consistent across multiple scales \cite{del2016discrete,del2021multiscale,dibak2018msm,hempel2021independent}.

In the context of data-driven modeling and simulation, a large range of methods have been recently constructed based on the Koopman operator  \cite{brunton2022modern,klus2016towards,klus2018data,korda2018linear,williams2015data}. One can intuitively construct the Koopman operator of system with fixed number of particles $n$ as follows: (i) the infinitesimal generator of the process is given by $\mathcal{A}_n$ from \cref{eq:generalstruct}, this could be the Liouville, the Fokker-Planck generator or something more complex ($\mathcal{Q}_n$ is not relevant for now as the particle number is fixed); (ii) the solution of the equation is given in terms of the exponential of the infinitesimal generator $\mathcal{A}_n$, which for a chosen time-step defines a propagator. This propagator can be understood as the Ruelle or Perron–Frobenius operator, which propagates probability densities forward in time; (iii) the adjoint of this operator is the Koopman operator and instead of propagating probability densities, it propagates observables. This last property is why it is suitable for data-driven methods. To the best of our knowledge, we are not aware of the construction or application of the Koopman operator for particle-based systems with varying copy numbers. This work delivers the infinitesimal generator for general dynamical systems with varying particle number, $\mathcal{A}_n+ \mathcal{Q}_n$ from \cref{eq:generalstruct}. This is the first cornerstone to derive the corresponding Koopman operator. Obtaining such an operator would enable applications in a diverse range of fields.

Many of these methods and techniques based on the general master equation can be applied to fields beyond physics. As the equation describes a general random dynamical system, it can be applied to complex systems with varying particle number beyond reaction-diffusion and molecular systems. This will potentially have significant applications in the development of numerical schemes and multiscale modeling of the spread of diseases \cite{britton2019epidemic,malysheva2022stochastic,winkelmann2021mathematical}, innovations \cite{djurdjevac2018human}, opinion dynamics \cite{castellano2009statistical,helfmann2023modelling}, and power, transportation and communication systems \cite{newman2018networks,helbing2001traffic} among others.

To finalize, we list some relevant examples of how the master equations presented in this work aid in the numerical simulation of complex molecular systems.
In particular in the treatment of multiscale molecular simulations that go beyond equilibrium, for example, systems in a temperature gradient \cite{noneq1}. Such a situation can be realized by embedding the open system in two distinct reservoirs, each at a different thermodynamic condition. In such a case the prescription for the simulator is reduced to the boundary conditions of the equation of \cref{sec:LiouvilleEqns} at the interface between the system and each reservoir. In general, the embedding of the open system in multiple reservoirs at different thermodynamic conditions allows the design of numerical algorithms, where the reservoir can be modeled as a thermodynamically fluctuating (in time) region via the fluctuating hydrodynamics method \cite{abbasprl}. The instantaneous thermodynamic condition of the reservoir corresponds to boundary conditions for the open systems which then it is simulated and whose averaged molecular properties are given, in the next step, as input to the reservoir; the procedure is then repeated thus producing the dynamics of the open system. Without the information on the boundary conditions of the equations of \cref{sec:LiouvilleEqns}, such a numerical scheme could not be implemented with such a physical consistency.  

In the realm of reaction-diffusion processes, the chemical diffusion master equation can be expanded in terms of classical creation and annihilation operators acting on the basis of the underlying space (a Fock space) \cite{del2022formulations,del2022probabilistic}. Based on this formulation, it is straightforward to develop Galerikin discretizations of the master equation \cite{del2022probabilistic}. This immediately yields the so-called reaction-diffusion master equations \cite{hellander2012reaction,winkelmann2016spatiotemporal}, where the space is discretized in voxels, and the diffusion is modeled as jumps between neighboring voxels. However, unlike previous standard derivations \cite{hellander2012reaction}, the scaling of the rates for nonlinear reactions (involving two or more particles) is automatically adjusted to the size of the lattice grid chosen, which enforces a consistent convergence to the micro and macroscopic scale \cite{del2022probabilistic,isaacson2013convergent,kostre2021coupling}. It further provides a relation between the microscopic parameters and the partial differential equation model at the macroscopic scale \cite{kostre2021coupling}, enabling particle-based simulations that are consistent with the macroscopic model and thus facilitating the implementation of reservoirs as macroscopic models that are coupled to a particle-based model \cite{kostre2021coupling}. 

The methodologies emerging from the theoretical constructs presented in this work do not only have the potential to provide physical insight, numerical schemes and solutions to the master equations of a large range of complex systems with varying particle number; they also have unifying capabilities and potential to bring insights into the underlying physics through the analysis of convergence properties, stability, and approximation errors. Exploring the application of these methods in diverse complex systems will allow researchers from several fields to uncover novel phenomena, unify models across scales and design efficient multiscale computational strategies for studying systems with varying number of particles.

\begin{acknowledgments}
	This work was partly supported by the Deutsche Forschungsgemeinschaft (DFG) Collaborative Research Center 1114 ``Scaling Cascades in Complex Systems'', project No.235221301, Project C01 (L.D.S.) ``Adaptive coupling of scales in molecular dynamics and beyond to fluid dynamics''. M.J.R. was supported by DFG grant no. RA 3601/1-1.
\end{acknowledgments}

\bibliographystyle{abbrv}
\bibliography{references}

\end{document}